\documentclass[useAMS,usenatbib]{mn2e}

\usepackage{epsfig}
\usepackage{longtable}
\usepackage{times}
\usepackage{color}
\usepackage{amsmath}

\def \mnras {MNRAS}
\def \apj {ApJ}

\newcommand{\msun}{{~M_{\odot}}}

\title[Episodic jet power and peak luminosity of soft state]{Discovery of the correlation between peak episodic jet power and X-ray peak luminosity of the soft state in black hole transients}
\author[Zhang \& Yu]{H. Zhang$^{1}$\thanks{E-mail: hzhang@shao.ac.cn} \& W. F. Yu$^{1}$\thanks{E-mail: wenfei@shao.ac.cn}  \\
$^{1}$Key Laboratory for Research in Galaxies and Cosmology,
Shanghai Astronomical Observatory, Chinese Academy of Sciences, 80
Nandan Road, Shanghai 200030, China}

\begin{document}

\date{}

\maketitle

\begin{abstract}
Episodic jets are usually observed in the intermediate state of
black hole transients during their X-ray outbursts. Here we report
the discovery of a strong positive correlation between the peak
radio power of the episodic jet $P_{\rm jet}$ and the corresponding
peak X-ray luminosity $L_{\rm x}$ of the soft state (in Eddington
units)  in a complete sample of the outbursts of black hole
transients observed during the RXTE era of which data are available,
which follows the relation $\log P_{\rm
jet}=(2.2\pm{0.3})+(1.6\pm0.2)\times \log {L_{\rm x}}$. The
transient ultra-luminous X-ray source in M31 and HLX-1 in EXO 243-49
fall on the relation if they contain stellar mass black hole and
either stellar mass black hole or intermediate mass black hole,
respectively. Besides, a significant correlation between the peak
power of the episodic jet and the rate-of-increase of the X-ray
luminosity $\rm dL_{x}/dt$ during the rising phase of those
outbursts is also found, following $\log P_{\rm
jet}=(2.0\pm{0.4})+(0.7\pm0.2)\times \log {\rm d}L_{\rm x}/{\rm
d}t$. In GX 339$-$4 and H 1743$-$322 in which data for two outbursts
are available, measurements of the peak radio power of the episodic
jet and the X-ray peak luminosity (and its rate-of-change) shows
similar positive correlations between outbursts, which demonstrate
the dominant role of accretion over black hole spin in generating
episodic jet power.  On the other hand, no significant difference is
seen among the systems with different measured black hole spin in
current sample. This implies that the power of the episodic jet is
strongly affected by non-stationary accretion instead of black hole
spin characterized primarily by the rate-of-change of the mass
accretion rate.

\end{abstract}


\section{Introduction}
Radio jet has been observed in Galactic black hole X-ray binaries
and neutron star X-ray binaries, in transient ultra-luminous X-ray
binaries in nearby galaxies which are suspected to contain stellar
mass compact objects or intermediate mass black holes (IMBHs), and
in active galactic nucleus (AGN) which contains supermassive black
holes (SMBHs). Accretion time scale in black hole X-ray binaries is
much shorter than that of the AGNs, which allows us to study black
hole accretion on longer accretion time scales unaccessible in AGNs
using current observations.  For instance, we are able to observe
distinct X-ray spectral states and spectral transitions in single
black hole X-ray binaries with X-ray all sky monitoring or pointed
observations. Distinct spectral states are probably exist in AGNs
\citep{kording2006}, but complete spectral state transitions have
not been seen due to limited observing time scale compared with
accretion time scales of SMBHs. Black hole X-ray binaries are
therefore unique sources to probe accretion physics such as spectral
state transitions.

Most black hole X-ray binaries (BHXBs) are transients
\citep{liu2007}. During outbursts, they will usually go through
several distinct X-ray spectral states, such as the hard state,
intermediate state (or steep power-law state) and the high soft
state \citep[see][for the definitions of these
  spectral states]{rm2006,done2007}. There are two kinds of radio jets
observed in BHXBs. One type is called continuous or
compact jet, which is observed in the hard state and switched off in
the high soft state. The continuous jets are compact and usually not
easy to resolve with current radio facilities, except in several cases, such as jets in GRS
1915+105 \citep{dhawan2000} and Cyg X-1 \citep{stirling2001}. The main
radiation component of the compact jet is thought to be self-absorbed synchrotron
emission \citep{markoff2001,yuan2005}, and usually shows flat or
slightly inverted spectrum in radio band. A correlation between the radio flux and the X-ray flux
$L_{Rad}\propto L_{X}^{\sim0.7}$) has been found in black
hole X-ray binaries \citep {hannikainen1998,corbel2000,corbel2003} and is
thought to be universal \citep{gallo2003}. However, some outliers and
an additional track have been found in recent years with much
different correlation index \citep[e.g.][]{soleri2010,coriat2011,ratti2012}.

The other type is called episodic or ballistic jet, which displays
intense radio flares during the intermediate state during the
transition between the hard state and the high soft state
\citep{fender2004}. At this time, the continuous jet has already
faded. The radio flares are associated with plasmoids ejections, and
sometimes can be resolved by radio observations, such as the jets
seen in GRS 1915+105 \citep{mirabel1994, fender1999}, GRO J1655-40
\citep{hjellmin1995,tingay1995}, XTE J1550-564 \citep{corbel2002}, H
1743-322 \citep{miller-jones2012} and XTE J1752-223
\citep{yang2010,yang2011}. The radio emission of such episodic jet
originates from an optically thin synchrotron emission component and
the Lorentz factor of the ejected plasmoids is usually much larger
than that of the continuous jet, and sometimes these episodic jets
appear superluminal (e.g. GRS 1915+105, \citealt{mirabel1994,
fender1999}; GRO J1655-40, \citealt{hjellmin1995,tingay1995}; XTE
J1748-288, \citealt{brocksopp2007}). The emission from the episodic
jet is probably more polarized than the continuous jet
\citep{fenderbelloni2004}.

Several models have been proposed for powering continuous and
episodic jets, from black hole spin \citep{blandford1977} to diverse
processes in the accretion flow
\citep{falcke1995,spada2001,markoff2001,yuan2005,yuan2009}. Whether
black hole spin is related to the jet power of both continuous or
episodic jet has become a central topic. For continuous jets,
\citet{fender2010} found no single relation between the black hole
spin and the jet power. This means that the compact continuous jets
are not powered by black hole spin. \citet{narayan2012} pointed out
that the compact continuous jet in the hard state was launched far
away from the radius \citep[$\sim 10$-$100
GM/c^{2}$,][]{markoff2005} where the effect of the black hole spin
is not important, so the spin has no significant effect on the
continuous jet power. However, the episodic jet was thought to be
launched within several gravitational radii and spin may play an
important role in powering the episodic jet. \citet{narayan2012}
studied the relation between the black hole spin estimated via X-ray
continuum-fitting method and the episodic jet power, which was
defined as the peak luminosity of the radio flare at 5 GHz. They
claimed that there is a positive correlation between the black hole
spin and the power of the episodic jet in five black hole X-ray
binaries, providing strong evidence that the episodic jets were
powered by the spin of the black hole. Later on, \citet{steiner2013}
followed the work of \citet{narayan2012} by adding one more black
hole X-ray binary in the data and further strengthened this
conclusion. However, \citet{russell2013} reached a different
conclusion. They have estimated the jet power with two different
ways, one way was to estimate the peak jet power by taking the peak
radio luminosity of the radio flare, the same as
\citet{narayan2012}, the other way was to estimate the minimum total
energy required to energize a synchrotron flare. They found there
was no relation between the jet power estimated as above and the
black hole spin estimated via either X-ray continuum method or X-ray
reflection method. Their results imply that the black hole spin does
not play an important role in powering the episodic jet. However,
\citet{mcclintock2013} pointed out the data selection in
\citet{russell2013} was not appropriate because they neglected
beaming effect.

For discrete/episodic ejections seen in earlier observations of
black hole binaries mostly in the transition phase (but in some
cases the radio peak may be associated with X-ray low/hard state,
see \citet{fender2001c}), a significant correlation between the peak
radio flux and the peak soft X-ray luminosity was found
\citep{fender2001a}. \citet{yu2004} pointed out there should be a
correlation between the peak hard X-ray flux (i.e., hard-to-soft
transition flux in most cases) and the peak radio flux in the hard
state because there is  a positive correlation between the hard
X-ray peak flux and soft X-ray peak flux for several low mass X-ray
binary transients. Such a correlation is expected if there is a
universal correlation between the radio flux and the hard X-ray flux
in the hard state\citep{gallo2003}. More studies on single sources
has shown that there is a correlation between the transition
luminosity and the peak luminosity of the following soft state in
single sources such as Aquila X$-$1 and GX 339$-$4
\citep{yu2007a,yu2007b}. Following that, a systematic study of
bright persistent and transient black hole and neutron star X-ray
binaries with data from the {\it RXTE}/ASM and the Swift/BAT by
\citet{yu2009} demonstrated that the correlation found by
\citet{yu2004} holds for both flares in persistent sources and
outbursts in transient sources. Furthermore, a correlation between
the rate-of-change of the luminosity around the hard-to-soft state
transition and the luminosity of the hard-to-soft state transition
was found as well. The series of studies indicate that
non-stationary accretion, represented by rate-of-change of the mass
accretion rate instead of the mass accretion rate alone, plays a
dominant role in the spectral state transitions and in producing the
outburst flux amplitudes. Motivated by the series of works,  it is
very important to study whether non-stationary accretion,
characterized by large rate-of-change in the mass accretion rate,
plays important role in generating episodic jet  in black hole X-ray
binaries during the transition from the hard state to the soft
state.

\section{Observations and Data Analysis}
We intended to obtain the peak jet power and the peak luminosity of
the soft state in the rising phase of any black hole transient
outburst that has been observed by the {\it RXTE}/ASM. We adopted
the peak luminosity of the radio jet at 5 GHz as the peak power of
the episodic jet following \citet{narayan2012}. The jet power is
estimated by using the following formula: $P_{\rm jet}=D^{2}(\nu
S_{\nu})_{\rm max,5GHz}/M$. This definition is actually
corresponding to the radio jet luminosity $L_{radio}$ which is
proportional to the jet power $Q_{jet}^{\xi}$( $\xi$ is in the range
1.22--1.58), when $Q_{jet}$ is a constant fraction of accretion
power $Q_{acc}$ \citep[see][]{falcke1995,heinz2003,coriat2011}. We
selected most of the sources from \citet{narayan2012},
\citet{steiner2013} and \citet{russell2013}, except GRS 1915+105 and
Cyg X-1. The X-ray observations of GRS 1915+105 allows us to search
for the soft state which followed radio flares. But we found that no
soft states can be associated with the radio flares in 1997, 2006
and 2010, based on the hardness ratio between {\it RXTE}/ASM and
{\it Swift}/BAT \citep[see the method in][]{yu2009} or the hardness
ratio calculated with the {\it RXTE}/ASM 3-band data. For the 1994
radio flare we did not have good soft X-ray coverage, so do not
allow us to determine the time range of the soft state. So we
excluded GRS 1915+105 in our plot, but we will discuss this source
in \S 4. We also exclude Cyg X-1, because the source didn't enter
the soft state after the radio flare on 2004 Feb
20\citep{fender2006}.

In addition, we have also analyzed radio observations of the black
hole transient \mbox{GX~339-4} and found evidence of an episodic jet
in the black hole transient \mbox{GX~339-4} during the intermediate
state of the 2010--2011 outburst, which has never been reported,
although the measurements of the radio fluxes during the hard state
of the outburst has been studied by \cite{corbel2013}. We processed
the radio observations of \mbox{GX~339-4} observed by Australia
Telescope Compact Array (ATCA) during its 2010--2011 outburst in
order to obtain the measurement of the peak flux of the episodic
jet. The ATCA synthesis telescope consists of six 22-m antennas
along the east-west direction. It was upgraded with the new Compact
Array Broadband Backend (CABB) system \citep{wilson2011}, and the
maximum bandwidth had been increased from the 128 MHz (in each of
two IF bands) to 2 GHz. This was coupled with higher level data
sampling, and has improved the continuum sensitivity significantly.
The observations of \mbox{GX~339-4} were carried out at the
frequencies of 5.5 and 9.0 GHz simultaneously. PKS 1934-638 was used
as the flux and bandpass calibrator, and the antenna's gain and
phase calibration, as well as the polarization leakage, were carried
out by nearby PKS 1740-517. We used MIRIAD software package
\citep{sault1998} to analyze the ATCA data. Our analysis of the
observations corresponding to the hard state has reproduced the
published results by \cite{corbel2013}. We then focused on the radio
data collected during its hard-to-soft transition in order to study
the episodic jet. We found an optically-thin radio flare during the
hard-to-soft state transition. The radio fluxes were averaged every
5 minutes for both bandpasses. We found that the radio flux
increased from $\sim$ 17 mJy to its peak flux at $\sim$ 23.5 mJy at
5.5 GHz in less than 4 hours. The radio emission was optical thin
during the entire observation. Figure 1 shows the ACTA light curve
of the radio flare event on April 27, 2010, and the averaged flux on
that day is shown on top right panel of Fig. 2 by the symbol of
magenta star. The peak radio flux of the episodic jet in the
outburst of 2010 is less than half of the peak radio flux of a
similar event in the 2002 outburst \citep{gallo2004}. The two
outbursts clearly show the proportionality that brighter outbursts
are associated with larger peak power of the episodic jets, implying
the peak power of the episodic jet is primarily determined, if not
entirely, by the accretion power that generates the X-ray outbursts.

In order to compare the peak power of episodic jet and the peak
luminosity of the soft state, we measured the corresponding peak
fluxes of the soft states in the long-term light curves obtained
with the {\it RXTE}/ASM (2-12 keV) for all the outbursts in black
hole X-ray binaries with the peak episodic jet power of black hole
transients reported before. Figure 2 shows the daily averaged ASM
light curves of the outbursts in nine black hole X-ray binaries in
which both measurements of the episodic jet power and the soft X-ray
peak are available. Data corresponding to the soft state are marked
in red based on previous reports. In half of these reports, the
definition of the soft state followed \citet{rm2006}, but in other
reports earlier on the sources GX 339-4, H 1743-322, XTE J1748-288,
XTE J1720-318 and XTE J1652-453 the definition of the soft state
could be inconsistent with the definition of \citet{rm2006}. For the
purpose to achieve consistent results, we have specifically studied
the power density spectra obtained with the {\it RXTE}/PCA
observations of those sources and measured the total
root-mean-square (rms) power integrated over 0.1 to 10 Hz, in order
to follow the same definition. The time ranges of the soft states we
derived from those RXTE observations are almost the same as those
reported we applied the rms threshold for the soft state as less
than 0.075 \citep{rm2006} and the differences are all within several
days. We found these will not bring significant difference in our
measurements of the peak flux of the soft state we selected based on
the soft state regime identified in previous reports, except for XTE
J1720-318, of which the time for the source to reach the peak flux
of the soft state was two days later than the one from the previous
report, but the flux difference can still be ignored due to larger
uncertainties in the mass and distance. So the peak fluxes we
measured are good estimates of the peak fluxes of the soft states.
The solid arrows mark the flux peaks of the soft states. There was
only one exception - we excluded the isolated ASM flux peak in the
soft state for XTE J1752-223 in order to avoid false measurement of
the peak flux, as discussed in \citep{yu2009}.

In order to accurately estimate the X-ray luminosity , we used the
{\it RXTE}/ASM data in the energy bands 1.3$\sim$3.0 keV,
3.0$\sim$5.0 keV and 5.0$\sim$12.2 keV. The Crab spectrum in 0.2-2
keV band, 2-10 keV band and 10-50 keV band can be described by a
power law with photon indices of -2.02, -2.07 and -2.12, and
normalizations of 8.95, 8.26 and 9.42, respectively
\citep{kirsch2005}.  We converted the count rates of the three bands
of {\it RXTE}/ASM (in Crab unit) into luminosities in 2-12 keV by
approximating these black hole X-ray binaries have similar spectral
shape and hydrogen absorptions to those of the Crab.
\citet{intzand2007} estimated that, if the photon index of the
actual source differs to the Crab index by 1 or $N_{H}$ is up to 10
times larger, the 2-12 keV energy flux differs by at most 30$\%$. We
have also simulated a disk-blackbody spectral component at 2 keV to
see if we can use a Crab power-law spectrum to fit the data and to
constrain the X-ray flux approximately in the three individual ASM
bands with XSPEC. We found that the deviations are 30\%, 4\% and 2\%
in the 1.3--3.0, 3.0--5.0 and 5.0--12.2 keV bands, respectively,
while most of the X-ray flux comes from the upper two energy bands.
So our estimates of the peak fluxes of the soft states are accurate
(and consistent) enough for the current study. The corresponding
peak luminosities (in Eddington unit,
$L_{Edd}=1.3\times10^{38}(M/\msun)$) of the high soft states of the
black hole X-ray binaries are listed in Table 1.

We have also managed to study the relation between the peak power of
the radio episodic jet and the rate-of-increase of the X-ray
luminosity during the rising phase in the hard state just before the
spectral transition from the hard state to the intermediate state.
We made use of the {\it RXTE}/ASM 3-band data to calculate the X-ray
flux over the time window from the 30\% of maximum ASM flux to the
ASM maximum flux in the hard state and fit the curves with a linear
relation, $L=b+k\times (t-50000)$. Here, L is the luminosity in 2-12
keV, t represents time in unit of Modified Julian Day, and $k$
represents the rate-of-change of the luminosity. The measurements of
the rate-of-change of the X-ray luminosity are also shown in Table
1. In Fig. 2, the data used in the fits are shown as green circle,
and the over-plotted blue solid lines correspond to the best-fits of
the slopes, but here the slopes represent the rate-of-change of the
X-ray flux.

\section{Results}
\subsection{Correlation between the episodic jet power and the peak luminosity of the soft state}
\subsubsection{Galactic black hole transients}

Figure 3 shows the relation between the power of the episodic jet
and the peak luminosity (in unit of Eddington luminosity) of the
high soft state for all the sources in which an episodic jet power
and the corresponding peak luminosity of the soft state can be
estimated with observations. Among the sources, we have relatively
good estimates of the distances and black hole masses for H
1743-322, XTE J1550-564, XTE J1752-223 and \mbox{GX~339-4} (The
corresponding references are shown in Table 1.), which are plotted
in black. Note that for XTE J1859+226, we only have the lower limit
of black hole mass, we also plotted in black since the uncertainty
in the mass would be within a factor of two. The distances and
masses are not well known for XTE J1652-453, XTE J1748-288 and XTE
J1720-318. The corresponding data are therefore plotted in green to
distinguish them from good measurements. In the plot, we have set
the distance to 8 kpc and the mass to 10 solar masses for both XTE
J1652-453 and XTE J1748-288 in the plot. For XTE J1720-318, our
knowledge of source distance and mass is even worse, so we set the
mass as 10 solar masses and the distance to be in the range from 3
kpc to 10 kpc, and the corresponding data are linked with a green
dashed line in Fig.~3. Most of the estimates of the mass and
distances were taken from previous works
\citep{narayan2012,steiner2013,russell2013}. In order to increase
our sample, we also looked into outbursts reported in A 0620-00, GS
1124-68, GS 2000+25 and GRO J1655-40 before the RXTE era. We
collected the peak fluxes in 0.4--10 keV from \citet{chen1997} (note
that whether the source are in the soft state is not clear). The
corresponding luminosities in Eddington unit were then estimated
with some updated source distances and masses (see Table 1). For the
outburst of GRO J1655-40 in 1994, the peak flux in 0.4--10 keV was
extrapolated from the reported peak flux in 20--100 keV
\citep{chen1997}, the source should have been in the hard state when
the hard X-ray flux reached its peak. So the X-ray luminosity in the
0.4--10 keV we estimated should be taken as the lower limit of the
peak flux of the following soft state. The radio activities of A
0620-00, GS 1124-68 and GS 2000+25 before the RXTE era were not well
covered, which does not  allow us to identify the radio flux peaks,
therefore the measurements of the jet power also correspond to the
lower limits. These samples are also plotted as green symbols in
Fig.~3 and Fig.~4.

When we only considered the sources with relatively good estimates
of distances and masses shown in black, which includes
\mbox{GX~339-4}, H1743-322, XTE J1550-564 and XTE J1752-223, and XTE
J1859+226, we found a strong correlation between the peak power of
the episodic jet and the corresponding peak luminosity of the soft
state. The Spearman rank correlation coefficient was 0.86, with the
chance possibility of 0.01. We also calculated the mean Spearman
rank correlation coefficient and its chance possibility when we
considered the uncertainties of the peak X-ray luminosity of soft
state and jet power using Bootstrap method. We found that the mean
Spearman rank correlation coefficient is 0.87$\pm$0.09, and the
possibility is 0.02$\pm$0.03, implying a strong correlation between
the peak power of the episodic jet and the peak luminosity of the
soft state. We fit the data established for the seven outbursts of
the above five sources with the linear function $\log P_{\rm
jet}=A+B\times \log L_{\rm soft,peak}$, and obtained the parameters
$A=2.2\pm0.3$ and $B=1.6\pm0.2$. The best-fit model is over-plotted
as a straight line in Figure 3.

We should point out that the results shown in Figure 3 include the
uncertainties estimated due to unknown distances or masses of black
hole X-ray binaries. Notice that the green data points are
systematically below those data shown in black except XTE J1748-288
and GRO J1655-40. This is consistent with the idea that the coverage
of the outbursts with radio observations which correspond to those
green data points was much worse than the coverage with the X-ray
observations due to the availability of X-ray monitors such as the
{\it RXTE}/ASM. Most of the radio flux peak corresponding to the
episodic jet in those sources marked in green were probably missed
due to sparse radio coverage. If we calculate the Spearman rank
correlation for all the black hole X-ray binaries except XTE
J1652-453 (which is put only a lower limit of the jet power) and set
the distance of XTE J1720-318 to 6.5 kpc (taking as the median of 3
kpc and 10 kpc - the range of previous measurements), we found the
correlation coefficient is 0.47$\pm$0.07 and the chance possibility
is 0.18$\pm$0.07. We found the source 4U 1543-475 has a significant
impact on the result since the coverage of radio observations of
this source was bad and the peak radio flux was probably missed (we
will discuss in next section). If we exclude this source, the
spearman correlation coefficient would be 0.70$\pm$0.06 and the
chance possibility would be 0.04$\pm$0.03.

\subsubsection{Extension to ultra-luminous X-ray sources in nearby galaxies}
The correlation between the power of the episodic jet and the
corresponding peak luminosity of the soft state (or the
rate-of-change of the X-ray luminosity) spans more than an order of
magnitude in the parameter space without showing a luminosity
saturation nor a cut-off at either ends of the X-ray luminosity
range. This characteristic suggests that more luminous or dimmer
episodic jet emission can be produced during more brighter or dimmer
outbursts respectively, which hints that episodic jet production is
somehow related to the occurrence of outbursts or specifically the
rate-of-change in the mass accretion rate. This is similar to the
conclusion obtained from the empirical relation between the
hard-to-soft transition luminosity and the peak luminosity of the
following soft state \citep{yu2009}. Both empirical correlations
probably extend to much higher luminosities such as those of the
ultra-luminous X-ray sources (ULXs) in nearby galaxies
\citep{yu2009}. On the other hand, as for the outbursts in the low
mass X-ray binary (LMXB) transients, no saturation of the peak
luminosity, up to the Eddington luminosity, is seen in the relation
between the X-ray outburst peak luminosity and the orbital period in
the Galactic transients \citep{wu2010}. These empirical relations
suggested that some stellar mass compact objects can turn into
ultra-luminous X-ray sources during brighter outbursts of similar
duration or shorter outbursts \citep{yu2009}, which has been indeed
observed \citep{middleton2013}.

The correlation between the peak power of episodic jet and the
corresponding peak luminosity of the soft state in Galactic sources
suggests that the investigation of the relation between the episodic
jet power and the peak luminosity of the soft state in transient
ULXs would tell us about the nature of the ULXs as compared with the
Galactic black hole transients; whether ULXs contains intermediate
mass black holes or just extreme versions of the Galactic
microquasars. In order to address this question, we have
investigated two transient ULXs in which radio emission from
episodic jet  has been detected during their outbursts. One of the
ULXs is XMMU J004243.6+412519, which was first detected by
XMM-Newton on 15 January 2012 \citep{henze2012} as an ultra-luminous
stellar-mass microquasar \citep[$\sim 10 \msun$,
  see][]{middleton2013} in M31, at a distance of only 0.78 Mpc away from
us. Bright radio emission was observed from this source with similar
behaviour to those Galactic black hole binaries
\citep{middleton2013}. The peak radio flux of the first VLA
observation ever measured was around 0.5 mJy at 5.26 GHz.  During
several additional radio flares for this source the peak radio flux
at 15 GHz was found around 1 mJy. So the peak radio flux at 5 GHz
can be estimated roughly in the range 0.5--1 mJy. The peak X-ray
luminosity of the soft state for the M31 ULX in 0.3--10 keV was
around 1.26 $\times 10^{39}$ erg s$^{-1}$ \citep{middleton2013}. We
plotted this source as magenta open squares in Fig.~4. It is obvious
that this source falls on the relation between the peak power of the
episodic jet and the peak luminosity of the soft state established
in the Galactic black hole transients as well. This implies that the
correlation seems to hold up to around the Eddington luminosities
for stellar mass black hole X-ray binaries.

The other ULX detected in radio is HLX-1. HLX-1 was first observed
by XMM-Newton on 23 November 2004 in the edge-on spiral galaxy ESO
243-49 about 95 Mpc away from us. The measured maximum X-ray
luminosity was about $1.1 \times 10^{42}$ erg s$^{-1}$, which
exceeds the Eddington luminosity of a stellar mass black hole by
about three orders of magnitude. Radio observations have been
performed with the Australia Telescope Compact Array (ATCA) during
two spectral state transitions of HLX-1, and radio flares were
detected in this source \citep{webb2012}. The peak flux density of
the radio flares for the combined 5 and 9 GHz data was 63 $\pm$ 18
$\mu$Jy. This value should be the lower limit of the flux density at
5 GHz given that similar radio flares observed in Galactic black
hole systems are usually optical thin. We therefore set this value
as the flux density at 5 GHz which should not deviate from the
actual value too much. In order to obtain the peak luminosity of the
soft state for HLX-1, we took the mass estimate of a black hole as
between $\sim 9 \times 10^{3} \msun$ and $\sim 9 \times 10^{4}
\msun$ \citep{webb2012}. We plotted these data of HLX-1 in Fig.~4 as
blue open circles with possible black hole mass of $9 \times 10^4
\msun$, $9 \times 10^3 \msun$ and 10 $\msun$, respectively (from
left to right in sequence). Interestingly, these data fall on the
correlation track formed from Galactic black hole binaries with good
mass and distance measurements. The data point corresponding to the
mass of $9 \times 10^3 \msun$ is the closest point to the
correlation track. This hints that HLX-1 is probably an intermediate
mass black hole if there exists a universal correlation between the
peak power of the episodic jet and the peak luminosity of the soft
state independent of black hole mass. However, we can not exclude
that HLX-1 may contain a stellar mass black hole considering the
95\% confidence interval of the fit in Fig. 4.

\subsection{Correlation between the peak power of episodic jet and the rate-of-change of the X-ray luminosity}
Figure 5 shows the relation between the power of the episodic jet
and the rate-of-change of the X-ray luminosity around the
hard-to-soft state transition measured from the ASM multi-band light
curves. When we only considered the data with good estimates of
X-ray flux shown in black, we found a very strong correlation. The
Spearman rank correlation coefficient is 0.91$\pm$0.06, and the
significance is 0.008$\pm$0.013. If we take the distance and black
hole mass of XTE J1748-288 as 8 kpc and 10 solar mass, respectively,
the Spearman rank correlation coefficient is 0.98 and the chance for
pure fluctuation is 3.3$\times 10^{-5}$. This indicates there exists
a strong correlation between the peak power of the episodic jet and
the rate-of-change of the X-ray luminosity in black hole transient
outbursts. We fitted the data in black and that of XTE J1748-288
with a linear function of the form $\log P_{\rm jet}=A+B\times \log
{\rm d}L_{\rm x}/{\rm d}t$, here the unit of ${\rm d}L_{\rm x}/{\rm
d}t$ is Eddington luminosity per day, we got $A=2.4\pm0.2$ and
$B=0.9\pm0.1$. If we excluded XTE J1748-288 and only consider the
data in black, we obtained $A=2.0\pm0.4$, and $B=0.7\pm0.2$. The
best-fit model is over-plotted in Fig. 5 in red.

\section{DISCUSSION}
\subsection{The accretion power and the episodic jet power}
Due to shorter accretion time scales and smaller distances, Galactic
black hole binaries are the best targets to study episodic jets as
compared with AGNs. Observations on time scales from days to years
allow us to discriminate between persistent compact jets and
episodic jets, which are usually seen in black hole hard state and
intermediate state, respectively. An important clue should be
noticed in Fig.~3. In the black hole binaries \mbox{H 1743$-$322}
and \mbox{GX~339$-$4}, the episodic jet power and the peak
luminosity of the soft state of the corresponding outburst have been
measured for two distinct outbursts. In either sources, a larger
power of the episodic jet during an outburst coincides with a larger
peak luminosity of the soft state, which clearly indicates that the
correlation is established in the accretion process among different
outbursts when the black hole mass and spin should stays almost the
same. Even if the total peak episodic jet power released might be
somehow related to black hole spin, the majority of the episodic jet
power should still come from the accretion power. In the current
data which does not allow sub-sampling on black hole spin, the
correlation established in outbursts of the same sources seems to be
consistent with the overall correlation established in outbursts of
our sample. This favors that a single physical process (parameter)
drives the correlation.

Since there is a correlation between the hard X-ray peak luminosity
(i.e., the luminosity of the hard-to-soft transition in outbursts
with state transitions) and the peak luminosity of the following
soft state \citep{yu2004,yu2007a,yu2007b,yu2009}, the correlation
between the peak episodic jet power and the peak luminosity of the
soft state in this study implies a correlation between the peak
luminosity of the hard state (or the luminosity of the hard-to-soft
state transition) and the peak power of the episodic jet.  This
seems to be supported by the apparent correlation between the power
of the episodic jet and the accretion power measured before episodic
jet ejection seen in GRS 1915+105. \citet{punsly2013} found that the
episodic jet power was strongly correlated with the intrinsic
1.2--50 keV X-ray luminosity 0-4 hours before the episodic jet
ejection, as well as the increase of the intrinsic X-ray luminosity
at 1.2--50 keV during the hours before the ejection in GRS 1915+105.

\subsection {Episodic jets under non-stationary accretion}
Episodic jets are usually seen during spectral transitions between
the hard state and the soft state when the source is in the
intermediate state or steep power-law state \citep{fender2004}. It
is known that there is a so-called `hysteresis' effect of spectral
state transitions during transient
outbursts\citep{miyamoto1995,nowak1995,maccarone2003}. The dramatic
hysteresis originate from the fact that the hard-to-soft spectral
state transition occurs in a large luminosity range by up to two
orders of magnitude in bright Galactic X-ray binaries
\citep{yu2004,yu2007a,yu2007b,yu2009}.  Systematic studies on the
hard-to-soft transitions in both persistent and transient bright
X-ray binaries show quantitative evidence that non-stationary
accretion, indicated by large rate-of-increase of the mass accretion
rate during the rising phase of transient outbursts, plays an
important role in determination of the luminosity at which the
hard-to-soft transition occurs as well as the outburst peak
luminosity (or peak luminosity of the following soft
state)\citep{yu2009}. In this work, we have found that there is a
correlation between the peak power of the episodic jet and the
corresponding peak luminosity of the soft state, as well as a
correlation between the peak power of the episodic jet and the
corresponding rate-of-increase of the X-ray luminosity in the 2-12
keV band during the rising phase of a transient outburst, indicating
that non-stationary accretion, represented by the observed large
rate-of-increase of the luminosity, appears to play an important
role in generating the episodic jet and the determination of its
peak power. We show that in multiple outbursts of a single source,
i.e GX339-4 and H1743-322, the larger the rate-of-increase of the
X-ray luminosity in the hard state is observed, the higher the peak
power of the episodic jet will be (see Fig. 5).

It is worth noting that the peak luminosity of the soft state and
the peak power of the episodic jet are not related to each other
linearly, thus the correlation cannot simply be explained as the
likely connection between the instantaneous mass accretion rate and
the mass loss rate into the jet. The relation between the peak
luminosity of the soft state and the peak power of the episodic jet
follows the relation $P_{\rm jet} \sim L_{\rm soft,peak}^{1.6}$.
Assuming the broadband spectrum of the episodic jet emission is of a
power-law with an index of -0.5 all the way up to the optical band,
the total jet luminosity is about 500 times the luminosity at 5 GHz,
while the total X-ray luminosity in 0.1-100 keV is about 12 times
the X-ray luminosity in 2-12 keV if we assume the photon index is
1.7. We found the prompt episodic jet luminosity will surpass the
X-ray luminosity when the X-ray luminosity in 0.1-100 keV is larger
than nine times Eddington luminosity. This gives an
order-of-magnitude estimate of the luminosity of the non-stationary
regimes corresponding to such a possible episodic jet-dominated
state. The luminosity limit also serves as a threshold to
discriminate whether the episodic jet power would primarily come
from an energy reservoir, e.g., the energy from the spinning black
hole by the Blandford-Znajek (BZ) effect or magnetic energy in the
accretion flow, or instead from the instantaneous accretion power.
Both the episodic jet-dominated state and the threshold are
interesting predictions by the empirical relations in the parameter
space not accessible now.

There are several ways to understand the relation between the peak
luminosity of the soft state and the peak power of the episodic jet,
which follows the relation $P_{\rm jet} \sim L_{\rm
soft,peak}^{1.6}$, when we recognize the role of the rate-of-change
of the mass accretion rate in interpreting the observations
unaccounted for by the mass accretion rate alone. One possibility is
that, if the radiation efficiency of the mass loaded in the episodic
jet does not change, then the ratio of the mass loss rate in the jet
to the mass accretion rate in the accretion flow should increase
with outburst amplitude and correlate with the rate-of-change of the
mass accretion rate. The typical fraction of the mass lose rate into
jet in those outbursts we studied as compared with the mass
accretion rate is probably only a few percent,  such as 0.5\%
\citep[see][]{yuan2005}. Then this mechanism can at most contribute
to a parameter space spanning by $\sim$ 200 times, from 0.5\% to
100\% of the mass accretion rate. The other possibility is that the
ratio of the mass loss rate in the jet to the mass accretion rate in
the accretion flow remains roughly constant but the radiation
efficiency of the episodic jet increases with peak luminosity of the
soft state and correlates with the rate-of-change of the mass
accretion rate. The limiting factor of this mechanism would be on
the order of 100 times as well, since the radiative efficiency of
jet is probably on the order of a few percent, such as 0.05
\citep{fender2001b,yuan2005} or 0.003 \citep{malzac2004}. Yet
another possibility is that the enhanced episodic jet power actually
comes from the black hole through B-Z effect, which requires the
extraction of black hole energy more efficient for sources with
brighter outbursts and for sources with higher spin. If this is
true, we would see systematic shifts of peak power of episodic jet
among black holes of different spins, forming parallel tracks of the
relation between the peak power of episodic jet and peak luminosity
of the soft state in Fig.3. This is yet not seen in the current
data, but an increase of the sample to several tens outbursts of
sources with diverse spins would allow meaningful constraints.

Spinning black hole as compared with non-spinning black holes would
have higher radiation efficiency for the emission of the the
episodic jet in the radio band as well as for the accretion flow in
X-rays by similar factors. This suggests that if we plot multiple
outbursts of individual black hole binaries in Fig.~3 or Fig.~4,
black hole binaries with different black hole spin would not bring
significant scattering around the overall empirical relation we
found, but only bring shifts up and down along the empirical
relation. If the power of the episodic jet primarily comes from the
energy extraction from the spinning black hole, we would observe
diverse correlations between the peak episodic jet power and the
peak luminosity of the soft state among sources with significantly
different black hole spins. This in turn suggests that towards the
higher luminosity end in the relation shown in Figure 3, a larger
scatter of observations data should exist if the episodic jet power
is strongly dependent on the black hole spin. Current observations
do not show such a sign.

\subsection{Episodic jets: early prediction of its peak radio flux}
\citet{yu2009} found the relation between the luminosity of
hard-to-soft state transition and the peak luminosity of the
following soft state in Eddington unit is in the form $\log L_{\rm
soft,peak}=(0.4\pm0.1)+(0.9\pm0.1) \log L_{\rm tr}$, where $L_{\rm
tr}$ represents the hard-to-soft state transition luminosity in
15-50 keV in unit of Eddington luminosity. So the relation between
the peak power of the episodic jet and the luminosity of
hard-to-soft state transition is derived as $\log P_{\rm
jet}=(2.9\pm0.4)+(1.5\pm0.3) \log L_{\rm tr}$. These empirical
relations derived above are useful. We are able to predict the peak
flux of the episodic jet based on the luminosity of the hard-to-soft
state transition and the rate-of-increase of the X-ray luminosity
before an episodic jet is launched. For example, when the luminosity
of the hard-to-soft state transition is only 0.001 $L_{\rm Edd}$ and
the black hole of 10 solar masses is at a distance of 8 kpc away
from us, we can roughly estimate that the peak flux of the episodic
jet at 5 GHz is $\sim$0.6 mJy. If the hard-to-soft transition occurs
at 0.1 $L_{\rm Edd}$, then the peak flux of the episodic jet at 5
GHz would be $\sim$ 0.7 Jy \citep[see][]{yu2009}. In some extreme
cases when the hard-to-soft transition occurs at nearly $\rm L_{\rm
Edd}$ and a distance of 0.78 Mpc (M31) or 3.25 Mpc (M82) for
ultraluminous X-ray sources, then the peak flux of the episodic jet
at 5 GHz would be $\sim$2.3 mJy or $\sim$0.1 mJy, respectively.
Normally episodic jet flares in microquasars are on the time scale
of hours, thus sensitive wide field-of-view radio facilities such as
Murchison Widefield Array (MWA) and Square Kilometer Array (SKA)
would be able to identify such extreme events in black hole X-ray
binaries in nearby galaxies and efficiently probe the nature of
ultra-luminous X-ray sources in nearby galaxies.

\section{Summary}
 In this paper, we have collected published radio data of episodic jets
observed in nine black hole X-ray binaries which were covered
simultaneously by the X-ray monitoring with the {\it RXTE}/ASM. In
addition, we have included our own detection of an episodic jet in
the ATCA radio observations of GX 339-4 in its 2010 outburst during
the hard-to-soft state transition. Using these data, we found a
strong correlation between the peak power of the episodic jet and
the peak X-ray luminosity of the soft state among the outbursts.
Data from multiple outbursts of individual sources and data among
all the sources seem to agree with a single relation, indicating the
power of the episodic jet is primarily accretion-driven. It is
surprising that two transient ULXs found in nearby galaxies in which
radio flares were detected during spectral state transitions also
fall on the same track. We also found a correlation between the peak
episodic jet power and the rate-of-increase of the X-ray luminosity
among the sources and among outbursts. These discoveries imply that
the power of the episodic jet is primarily driven by non-stationary
accretion during the rising phase of transient outbursts, the scale
of which varies from outburst to outburst and from source to source.

Our results not only confirm the previous discovery of a positive
correlation between the peak X-ray flux of X-ray outburst and the
peak radio flux density for black hole binaries \citep{fender2001a},
but also reveal the importance of rate-of-change in the mass
accretion rate to the episodic jet power. In this paper we intend to
focus on the origin of the power of the episodic jet in black hole
transients. We found two correlations, namely the correlation
between the peak power of the episodic jet and the peak luminosity
of the soft state, and the correlation between the peak power of the
episodic jet and the rate-of-increase of the X-ray luminosity around
the hard-to-soft state. Our results provide strong evidence that
non-stationary accretion is the driving source of the episodic jets
seen in black hole transients. Models about jet production should
consider the influence of the apparent rate-of-change of the mass
accretion rate shown in the change of the X-ray flux.

\section*{Acknowledgments}

We thank the anonymous referee's comments, and we also thank
Jean-Pirre Lasota, Feng Yuan and Qinwen Wu for their good
suggestions and comments. We thank the {\it RXTE} and the {\it
Swift} Guest Observer Facilities at NASA Goddard Space Flight Center
for providing the {\it RXTE}/ASM products and the {\it Swift}/BAT
transient monitoring results. This work was supported in part by the
National Natural Science Foundation of China (245681001, 11333005,
11073043, and 11350110498) and by Strategic Priority Research
Program "The Emergence of Cosmological Structures" under Grant No.
XDB09000000 and the XTP project under Grant No. XDA04060604, by the
Shanghai Astronomical Observatory Key Project. This work has made
use of the data obtained through the High Energy Astrophysics
Science Archive Research Center Online Service, provided by the
NASA/Goddard Space Flight Center. This paper also includes analysis
of archival data obtained through the Australia Telescope Online
Archive (http://atoa.atnf.csiro.au).

\clearpage

\begin{figure*}
\begin{center}
\centerline{\includegraphics[width=14cm,angle=0]{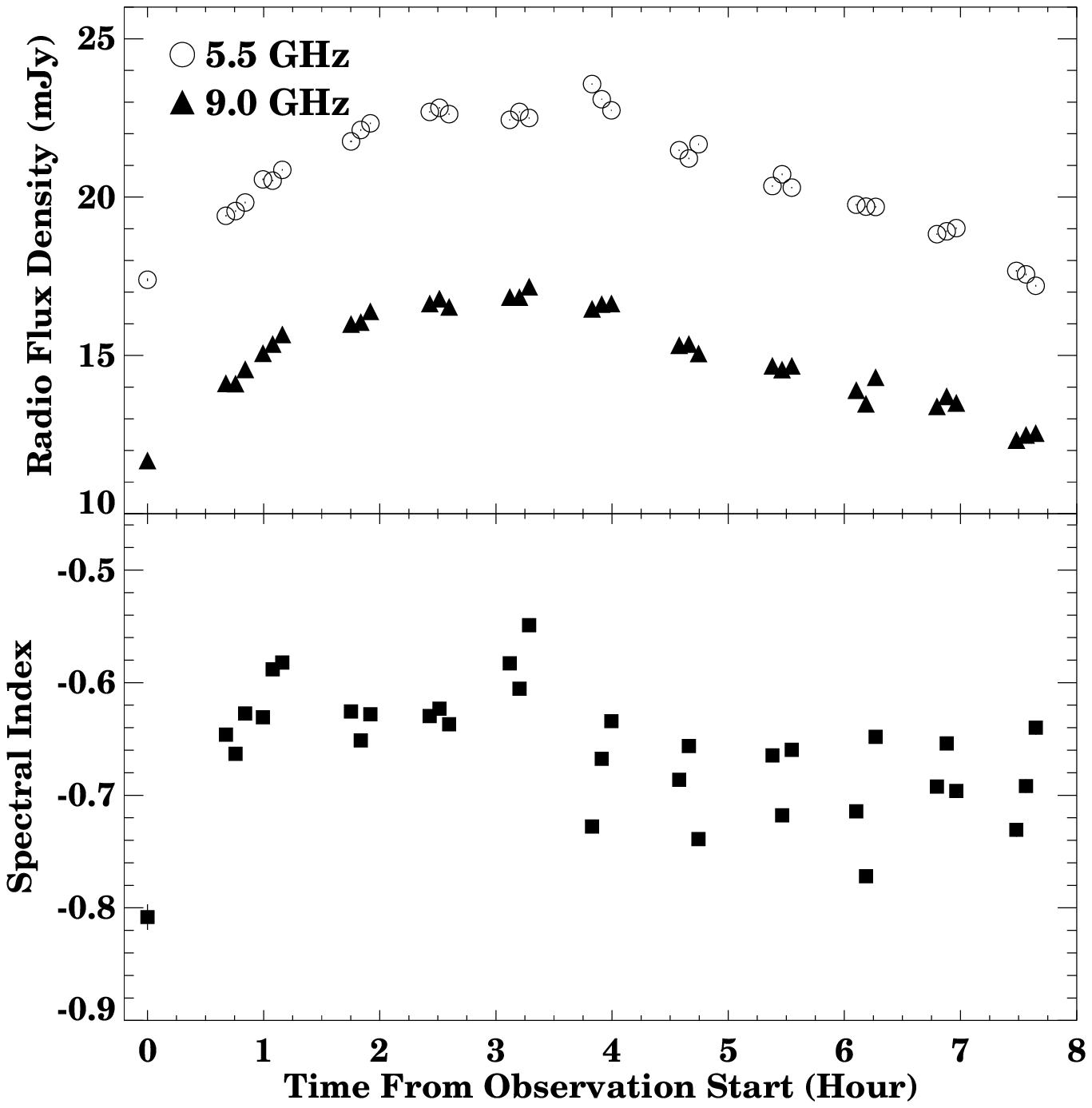}}
\caption{The radio flare of GX
339-4 seen with the ATCA on 2010 April 27 (MJD 55 313) at
5.5 and 9.0 GHz (upper) and the corresponding spectral index $\alpha$ (lower, where
$S_{\nu}\propto \nu^{\alpha}$). The error bars are overlapped by the
symbols. The radio fluxes are averaged over 5 minutes. }
\end{center}
\end{figure*}

\begin{figure*}
\begin{center}
\centerline{\includegraphics[width=18cm,angle=0]{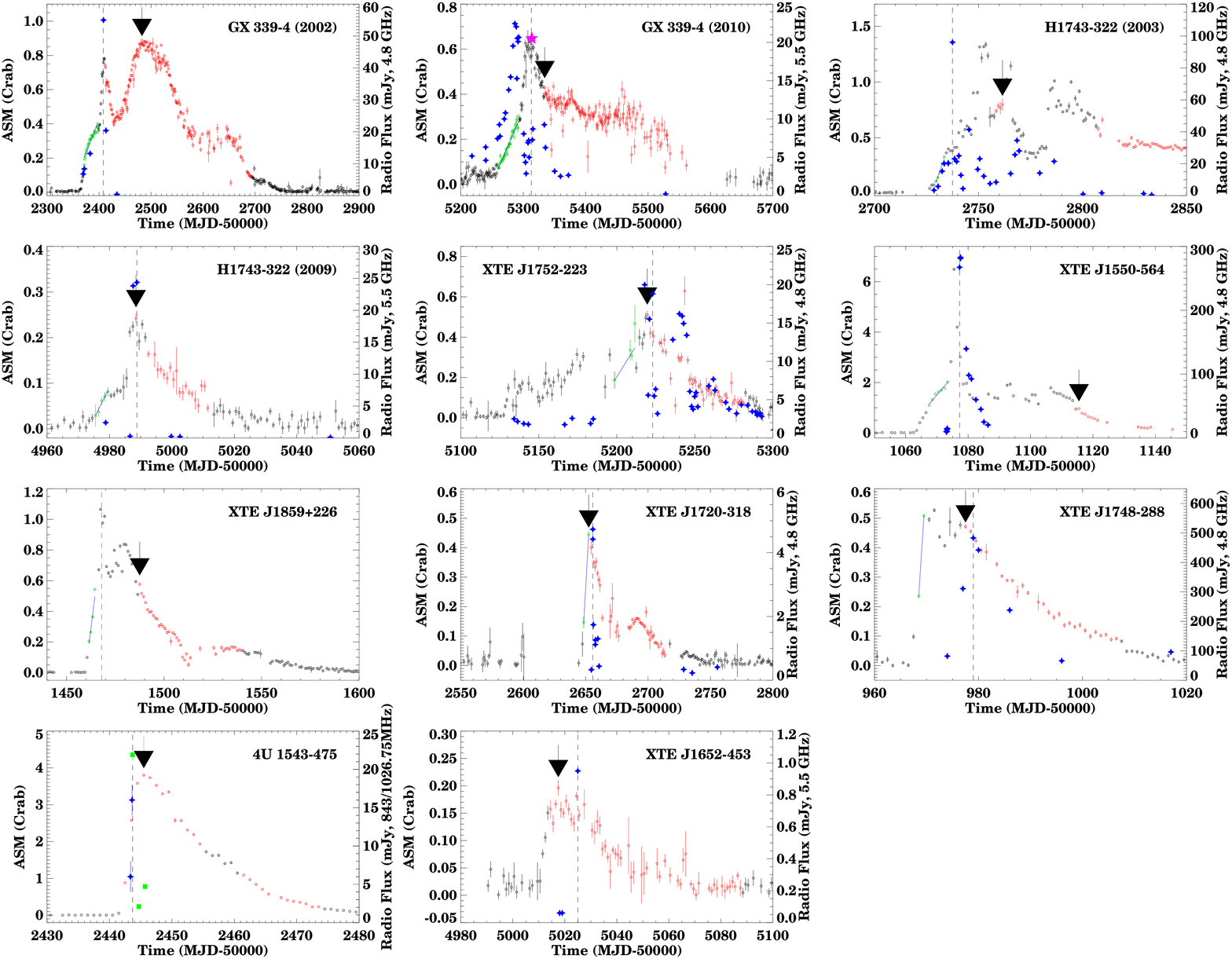}}
\caption{X-ray and radio observations of a sample of outbursts seen
with the {\it RXTE}/ASM in the RXTE era. The vertical dashed line
shows the time of the radio peak flux from the episodic jet. The
blue stars and green filled squares represent flux measurements from
radio observations, the magenta stars in top middle panel represents
the peak of the radio flare event of GX339-4 in 2010. The radio
light curve of XTE J1859+226 can refer to \citet{brocksopp2002}. The
red filled circles represent data corresponding to the high soft
state, for GX 339-4
\citep{homan2005,smith2003,motta2010a,motta2010b,dincer2012},
H1743-322 \citep{mcclintock2009, miller-jones2012}, XTE J1752-223
\citep{nakahira2012}, XTE J1550-564 \citep{sobczak2000}, XTE
J1859+226 \citep{rm2006}, XTE J1720-318 \citep{brocksopp2005}, XTE
J1748-288 \citep{brocksopp2007,revnivtsev2000} and 4U 1543-475
\citep{park2004}. Please note that the intermediate state could be
brighter than the soft state in several black hole transient
outbursts in 2-12 keV band, which cover both thermal disk component
and power-law component. The data marked as green circles were fit
to a straight line (shown in blue) to measure its slope and to
derive the rate-of-increase of the X-ray luminosity in hard state.
For 4U1543-475 and XTE J1652-453, no reliable data are available to
measure the rate-of-change of X-ray luminosity. Since radio emission
from the episodic jets comes from ejected blobs,  in a few cases the
radio flux peak is found when the source had entered the soft
state.}
\end{center}
\end{figure*}

\begin{figure*}
\begin{center}
\centerline{\includegraphics[width=14cm,angle=0]{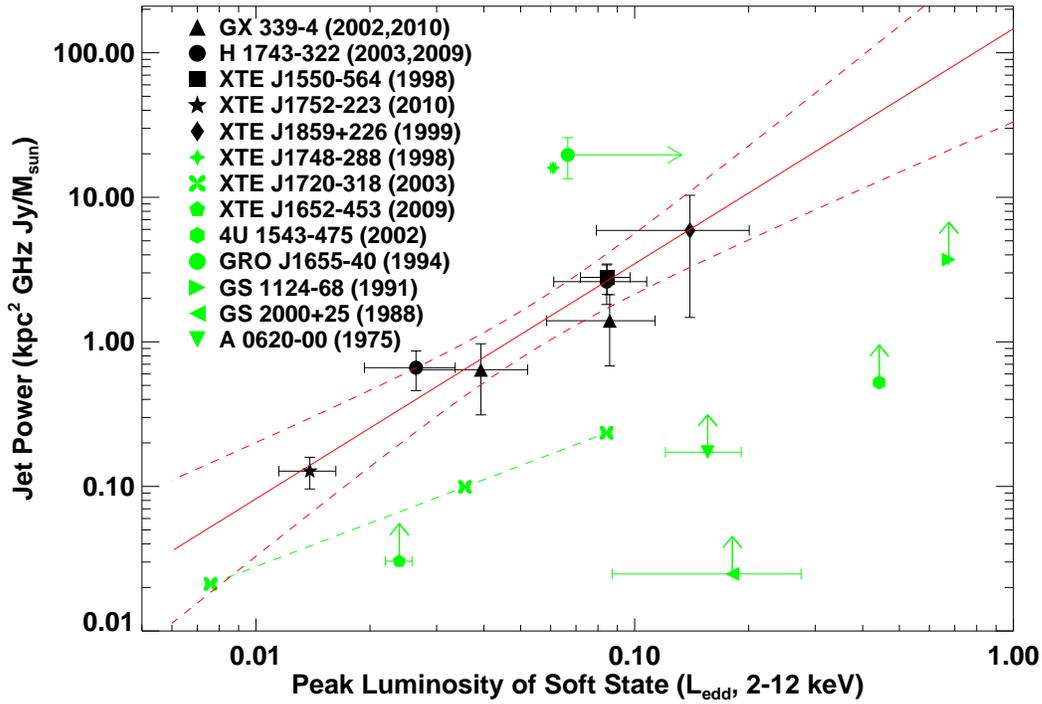}}
\caption{The relation between the episodic jet power and the peak
luminosity of the soft state in black hole transients. The black
points correspond to sources have good estimates of mass, distances,
X-ray and radio fluxes. The only exception is XTE J1859+226, which
has only a lower limit of black hole mass. The green data points
correspond to sources with a lot larger uncertainties in the
estimates of their mass or distances. We show XTE J1720-318 with
possible distances of 3, 6.5 and 10 kpc and the corresponding data
points are connected with a green dashed line. Data corresponding to
two outbursts of GX 339-4 and H 1743-322 demonstrate the dominate
role of accretion in producing radio jet power. For the sources A
0620-00, GS 1124-68, GS 2000+25 and GRO J1655-40 in which radio
flares were reported before the RXTE era, the peak X-ray luminosity
are calculated based on the extrapolations to the 0.4--10 keV band
from hard state hard X-ray data, and the radio fluxes should
correspond to the lower limits on the peak radio jet power due to
very sparse coverage at early times. The red solid line are the
best-fit for the black points in logarithm scale, and the red dashed
lines are the 95\% confidence intervals of the fit. The error bars
of each data points are primarily from the uncertainties in the
estimates of their distance and masses.}
\end{center}
\end{figure*}

\begin{figure*}
\begin{center}
\centerline{\includegraphics[width=14cm,angle=0]{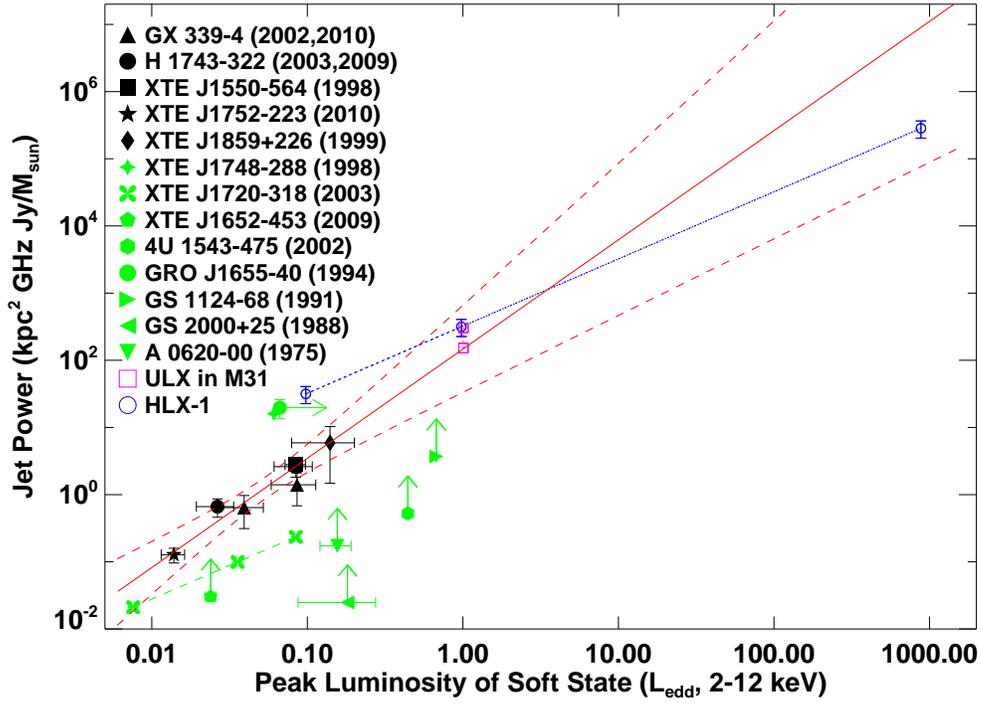}}
\caption{The relation between the episodic jet power and the peak
luminosity of the soft state in black hole transients and a few
ULXs. We add the data of HLX-1 and the ULX XMMU J004243.6+412519 in
M31 to illustrate that the ULXs in nearby galaxies are also
consistent with the relation found in Galactic black hole X-ray
binaries. The blue circles which are connected with blue dotted line
corresponds to different black hole mass estimates of HLX-1. From
left to right, we took the black hole mass as $9 \times 10^4 \msun$,
$9 \times 10^3 \msun$ and 10 $\msun$. The magenta open squares
correspond to radio measurements of 0.5 and 1 mJy for the ULX in
M31. The red solid and dashed lines are the same as that in Fig. 4.}
\end{center}
\end{figure*}

\begin{figure*}
\begin{center}
\centerline{\includegraphics[width=14cm,angle=0]{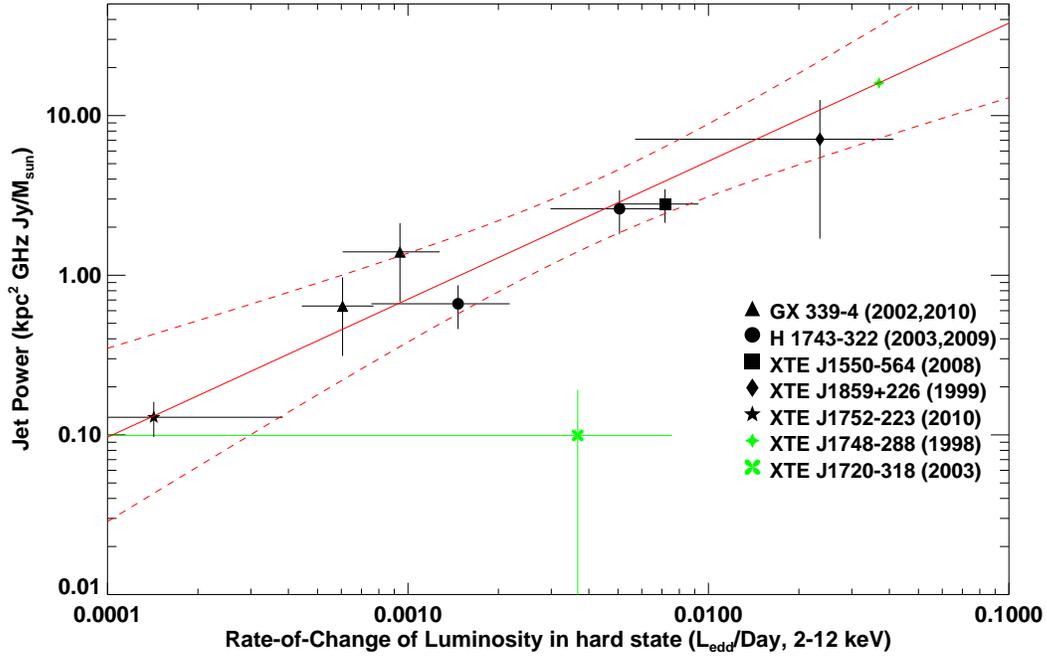}}
\caption{The relation between the peak power of the episodic jet and
the rate-of-change of the X-ray luminosity in the rising phase of
the outbursts. The data shown in black correspond to sources with
rather definite ranges of masses, distances, X-ray and radio fluxes,
except for XTE J1859+226 which has only a lower limit of black hole
mass. The green data points indicate sources a lot more larger
uncertainty in the estimates of their masses and distances. In the
plot, we set the distance to 6.5 kpc for XTE J1720-318. The red
solid line are the linear fit for the data in black in logarithm
scale, and the red dashed lines are the 95\% confidence intervals of
the fit.}
\end{center}
\end{figure*}

\clearpage

\begin{table*}
 \centering
\caption{List of the sources and parameters used}
\begin{tabular}{llllcclccc}
\hline \hline
Source &d  & Mass & $S_{\rm 5GHz}$&L$_{\rm max}$/L$_{\rm Edd}$&$dL/dt$&Year&\multicolumn{3}{c}{ References }    \\
&($\rm kpc$) &($\msun$)  &(mJy) &(\%) &(10$^{-2}$ L$_{\rm Edd}$/Day)& &d &Mass&$S_{\rm 5GHz}$\\

\hline
4U 1543-475                      &7.5 $\pm$ 1.0  &9.4 $\pm$ 1.0 &17.5           &44.2$\pm$8.4   &$-$ &2002   &1  &1  &2   \\
GX 339-4                         &   8$\pm$2     &12.3$\pm$1.4  &$\sim$ 55      &8.6$\pm$2.7    &0.09$\pm$0.03 &2002   &3  &4  &5\\
                                 &               &              &$\sim$24.5     &3.9$\pm$1.3    &0.06$\pm$0.02 &2010   &   &   & \\
H 1743-322                       &8.5$\pm$0.8    &13.3$\pm$3.2  &96.1           &8.4$\pm$2.3    &0.51$\pm$0.21 &2003   &6  &4  &7 \\
                                 &               &              &24.4           &2.6$\pm$0.7    &0.15$\pm$0.07 &2009   &   &   &8 \\
XTE J1550-564                    &4.38$\pm$0.50  &9.10$\pm$0.61 &265.0          &8.5$\pm$1.3    &0.72$\pm$0.21 &1998   &9  &9  &10,11\\
XTE J1652-453$^{(a)}$   &   8           &10            &$>$0.95                 &2.4$\pm$0.2    &$-$ &2009   &--$^a$   &--$^a$   &12\\
XTE J1720-318$^{(b)}$   &3$\sim$10      &10            &4.7             &3.6$\pm$0.1$^{(c)}$    &0.37$\pm$0.39 &2003   &13 &--$^b$ &14\\
XTE J1748-288$^{(a)}$   &8              &10            &500                     &6.1$\pm$0.1    &3.69$\pm$0.11 &1998   &--$^a$   &--$^a$   &15\\
XTE J1752-223                    &3.5 $\pm$ 0.4  &9.6 $\pm$ 0.9 &20             &1.4$\pm$0.2    &0.01$\pm$0.02 &2010   &16 &16 &17\\
XTE J1859+226                    &8$\pm$3        &$>$5.42       &100            &14.0$\pm$6.1   &2.35$\pm$1.78 &1999   &18 &19 &20  \\
GRO J1655-40                     &3.2$\pm$0.5    &6.30$\pm$0.27 &2420$^{(d)}$         &$>$6.7    &$-$   &1994   &21 &22 &21 \\
A 0620-00                        &1.06$\pm$0.12  &6.61$\pm$0.25 &$>$203$^{(d)}$       &15.6$\pm$3.5   &$-$   &1975   &23 &23 &24  \\
GS 1124-68                       &5.0            &6.0$\pm$1.5   &$>$171$^{(d)}$       &67.5$\pm$0.0   &$-$   &1991   &25 &26 &27  \\
GS 2000+25                       &2.7$\pm$0.7    &7.2$\pm$1.7   &$>$4.9$^{(d)}$       &18.1$\pm$9.4   &$-$   &1988   &28 &28 &29    \\

\hline
\end{tabular}

$^{(a)}$assume a distance of 8 kpc and a black hole mass of 10
$\msun$. $^{(b)}$assume a distance of 3 kpc and 10 kpc,
respectively, and a black hole mass of 10 $\msun$. $^{(c)}$ assume a
distance of 6.5 kpc.
$^{(d)}$measured in 0.4--10 keV.\\
(1)\citet{orosz2002}; (2)\citet{park2004}; (3)\citet{zdziarski2004};
(4)\citet{shaposhnikov2009}; (5)\citet{gallo2004};
(6)\citet{steiner2012}; (7)\citet{mcclintock2009}; (8)
\citet{miller-jones2012}; (9)\citet{orosz2011};
(10)\citet{hannikainen2009}; (11)\citet{narayan2012};
(12)\citet{calvelo2009}; (13)\citet{chaty2006}
;(14)\citet{brocksopp2005}; (15)\citet{brocksopp2007};
(16)\citet{shaposhnikov2010}; (17)\citet{brocksopp2013};
(18)\citet{hynes2005}; (19)\citet{corral-santana2011};
(20)\citet{brocksopp2002};(21)\citet{hjellmin1995};
(22)\citet{greene2001};(23)\citet{cantrell2010};
(24)\citet{kuulkers1999};(25)\citet{gelino2001};
(26)\citet{rm2006};(27)\citet{ball1995};
(28)\citet{barret1996};(29)\citet{hjellming1988}.
\end{table*}

\clearpage

\end{document}